\documentclass[preprintnumbers,amsmath,amssymb,onecolumn]{revtex4}
\usepackage[colorlinks=true,citecolor=red,filecolor=green,linkcolor=blue,pdfnewwindow=true]{hyperref}
\usepackage{graphicx}
\usepackage[title]{appendix}
\makeatletter

\renewcommand*{\p@subsection}{}

\renewcommand*{\p@subsubsection}{}
\makeatother
\begin{document}
\title{Non-Markovian process with variable memory functions}

\author{Athokpam Langlen Chanu, Jyoti Bhadana and R.K. Brojen Singh}
\email{brojen@jnu.ac.in (corresponding author)}
\affiliation{School of Computational \& Integrative Sciences, Jawaharlal Nehru University, New Delhi-110067, India.}

\begin{abstract} 
{\noindent}We present a treatment of non-Markovian character of memory by incorporating different forms of Mittag-Leffler (ML) function, which generally arises in the solution of fractional master equation, as different memory functions in the Generalized Kolmogorov-Feller Equation (GKFE). The cross-over from the short time (stretched exponential) to long time (inverse power law) approximations of the ML function incorporated in the GKFE is proven. We have found that the GKFE solutions are the same for negative exponential and for upto frst order expansion of stretched exponential function for very small $\tau \rightarrow 0$. A generalized integro-differential equation form of the GKFE along with an asymptotic case is provided. 
\end{abstract}

\maketitle

\section{Introduction}
\noindent Stochastic treatment in complex systems study has been a very powerful technique to study inherent properties in such systems because modelling with this technique, generally, closely mimics with actual experimental situations and can reveal dynamical patterns and behaviors, especially the importance of noise in the system dynamics. Kolmogorov-Feller (KF) equation in stochastic theory is an integro-differential equation governing the transition probability density (field) of stochastic processes with jump changes in the state of the system under study \cite{kolmo,feller1,Feller}, and has been applied to various areas of immense interest. In \cite{exact}, the authors have produced the exact stationary solutions to the classical KF equation for some examples of probability distribution analytically. The probability density function of bounded jump processes having saturation function and Poisson white noise is deduced and its analytical stationary solutions are found \cite{denisov}. KF equation is also used in some topics of astrophysics \cite{petro1, petro2, galaxy, comet}. This classical KF equation is extended to generalized Kolmogorov-Feller equation for one and two dimensional cases in operator representation and is solved using successive approximations methods\cite{zas}. The treatment of fractional generalization of the KF equation is done and a solution for the asymptotic time case is provided \cite{saichev}. \\ \\
\noindent The Kolmogorov-Feller equation is Markovian or memoryless in character. However, any natural dynamical process is not truly Markovian. Non-Markovian character, may be in the form of time delay, exists everywhere whether big or small. This necessitates a revisit to the classical Markovian Kolmogorov-Feller equation. In \cite{volatility,iaeng,stochastic}, the classical KF equation is analyzed to reach a generalized form of the KF equation with the treatment of memory function in the equation. The Generalised Kolmogorov Feller Equation (GKFE) is a very important formalism to study stochastic volatility, for instance in financial system such as financial time series analysis of stock market  \cite{iaeng,volatility}. However, the authors in \cite{volatility,iaeng,stochastic}, even though they study computationally the solution of GKFE, have not properly discusssed how they take, for a specific $m(t)$ memory function, the category of functions of $n(t)$ form. In the present paper, we attempt to use different forms of memory function and analyse the solution of GKFE analytically. We will discuss how to find, for an $m(t)$ memory function, the type $n(t)$ particularly the Mittag-Leffler function.\\ \\
\noindent Mittag-Leffler (ML) function, discovered by Swedish Mathematician, G$\ddot{o}$sta Mittag-Leffler in 1903 \cite{ml,erdel}, is a very important function in fractional calculus and rightly, known as fractional calculus's \textit{Queen} function. This functional form arises generally in the solution of master equation which has a fractional form, like, generalized  fractional kinetic equation, the study of super diffusive transport, studies related to random walk and L$\acute{e}$vy flights and also many complex systems \cite{saxena} study. During recent years, ML function has been studied in several applied problems of physics, chemistry, biology, engineering, earth sciences, etc. like electrical networks, fluid dynamics, rheology, diffusion transport and also in theories of probability and statistical distribution. The function ML is used in the fractional modeling of some stochastic processes like the model of Continuous Time Random Walk (CTRW) in finance \cite{mainardi, gorenflo}. This fractional calculus has been used to study properties of viscoelastic materials since it is able to model phenomena with longer memory (hereditary) \cite{vis1, vis2, vis3}. Kenneth S. Cole (1933) can be quoted for first applying ML function in connection with nerve conduction; the solution to the fractional order differential equations has an exponential functional generalized to the function ML and to the experimental data regarding cell membrane, this gives a better fit \cite{vis1,role}. Recently, ML function and its properties have also been studied in neuroscience \cite{neuro1, neuro2, neuro3} and cosmology\cite{cosmo, cosmo1}. In the present paper, we will discuss different approximations of Mittag-Leffler function and why we can use this function as a memory function for our GKFE. \\ \\
\noindent Our work is organized in the following way. Section 1 provides a brief introduction to Classical Kolmogorov-Feller Equation, Generalized Kolmogorov-Feller Equation and Mittag-Leffler function. Section 2 deals with the Classical Kolmogorov-Feller equation which is derived from the Brownian motion and kinetic theory of gas. Generalized Kolmogorov-Feller equation with memory function is also presented in section 2. Section 3 discusses Mittag-Leffler function, followed by two common asymptotic approximations of the ML function given in section 3.1. Section 4 is concerned with results of the analytical solutions of GKFE using different memory functions. Discussions based on the results obtained are given in section 5. Section 6 presents conclusion.\\
\section{Theory of Kolmogorov-Feller equation}
\noindent As early as 1905, Albert Einstein has used stochastic modelling of a natural dynamical process known as Brownian motion \cite{albert} which is a kind of random walk process, based on the idea of Markov chain developed around the same time by A.A. Markov \cite{markov}. In \cite{albert}, Einstein has studied the disordered Brownian motion of particles in a liquid arising out of thermal molecular motion that leads to diffusion. In such motion, one can consider a very small time interval $\tau$ in comparison to the observable time intervals. An important assumption in this study is that the motions carried out by a single particle during any two adjacent intervals of time are taken as independent events. Consider that $N$ total number of Brownian particles is inside the liquid having volume $V$ such that density function is denoted by $\mathcal{U}(t,x)$. Although the measured value per unit volume can be different, the distribution of the particles will almost remain the same throughout the volume. Thus, using Markov property, one can determine what is the particle distribution at some time $(t+\tau)$ from the knowledge of its previous distribution at time, $t$. In one-dimensional scenario with position $x$, we can thus write the particle number at any $x$ at time $(t+\tau)$ as \cite{albert},
\begin{equation}
\mathcal{U}(t+\tau,x) = \int \limits_{\delta=-\infty}^{\delta=+\infty} \ \mathcal{U}(t,x-\delta) \ f(\delta) \ d \delta \label{eq:1}
\end{equation}
We can easily observe the Markov assumption in equation (\ref{eq:1}). The function, $f(\delta)$ is any probability density function (p.d.f) for one-dimensional distance $\delta$ traversed by the particle during $\tau$ time interval. One can choose $f(\delta)$ depending on the properties and behavior of the system dynamics, for instance, a Gaussian p.d.f. The definition of convolution of any two functions, $G(a)$ and $H(a)$ gives,
\begin{eqnarray}
G(a)\otimes_a H(a)=\int \limits_0^a G(z) \ H(a-z)\ dz=\int \limits_0^a G(a-z)\ H(z) \ dz\nonumber
\end{eqnarray}
Using the above definition and rewriting equation \eqref{eq:1}, one can arrive at \cite{volatility,iaeng},
\begin{equation}
\mathcal{U}(t+\tau,x)= \mathcal{U}(t,x) \otimes_x f(x) \label{eq:2}
\end{equation} 
Here, $\mathcal{U}(t+\tau,x)$ can be regarded as a measured value given by expectation value of $\mathcal{U}(t,x)$ over the range of $x-$values. If we consider a Markovian (memoryless) process in the time evolution of the system, such that $\tau$ is negligibly small ($\tau \ll1$), one can Taylor expand R.H.S. of equation \eqref{eq:2} in powers of $\tau$ and take only up to first order term as,
\begin{eqnarray}
\label{eq:x}
\mathcal{U}(t+\tau,x)\approx \mathcal{U}(t,x)+\tau \frac{\partial  \mathcal{U}(t,x)}{\partial t}
\end{eqnarray}
From equations (\ref{eq:2}) and (\ref{eq:x}),
\begin{eqnarray}
\label{eq:4}
\tau \frac{\partial \mathcal{U}(t,x)}{\partial t} &=& -\mathcal{U}(t,x)+ \mathcal{U}(t,x)\otimes_x f(x) \label{rr}
\end{eqnarray}
This evolution equation of the field is generally known as the \textit{Classical Kolmogorov-Feller (KF) Equation} \cite{volatility}. It represents a Markovian process of the system with no memory. It assumes that only events with short term memory affect the field's evolution (dynamics) whereas long term memory (which depends on histories) events do not affect the behavior of the field. \\ \\
Around 1930s, Kolmogorov and Feller generalized the transition probability of Markov stochastic processes with jump changes in the state of the system under study \cite{kolmo,feller1}. Thus, for a stochastic jump process continuous in time, Kolmogorov \cite{kolmo} wrote the integro-differential equation of the Markov function $F(t,x)$ as,
\begin{equation}
   \frac{\partial F(t,x)}{\partial t}=-\int \limits_{-\infty}^{\infty} \ F(t,x) \ G(t,x,z)\ dz +\ \int \limits_{-\infty}^{\infty}F(t,z) \ G(t,x,z)\ dz \nonumber
\end{equation}
According to Feller \cite{feller1}, the forward and backward integro-differential equations are respectively, 
      \begin{equation}
     \frac{\partial P(\tau,x;t,\Lambda)}{\partial t}=-\int \limits_\Lambda p(t,y) P(\tau,x;t,dE_y)+\int \limits_E p(t,y) \Pi(t,y,\Lambda) P(\tau,x;t,dE_y)\nonumber 
      \end{equation}
      and
      \begin{equation} \frac{\partial P(\tau,x;t,\Lambda)}{\partial \tau}=p(\tau,x)[ P(\tau,x;t,\Lambda)-\int \limits_E P(\tau,y;t,\Lambda) \Pi(\tau,x,dE_y)] \nonumber 
      \end{equation}
\noindent Now, in equation \eqref{rr}, the investigation of the behavior of $\mathcal{U}(t,x)$ due to longer term events can be done by incorporating the role of a \textit{memory function}, $m(t)$ in the KF equation (\ref{eq:4}) as given in \cite{volatility,iaeng} as
\begin{eqnarray}
\label{eq:5}
m(t) \otimes_t \frac{\partial \mathcal{U}(t,x)}{\partial t}=-\frac{\mathcal{U}(t,x)}{\tau} +\frac{1}{\tau} \  \mathcal{U}(t,x)\otimes_x f(x)
\end{eqnarray}
If one takes $m(t)=\delta(t)$, because of the convolution property of the delta function $\delta \otimes h= h$, R.H.S. in the equation (\ref{eq:5}) becomes,  
$\delta(t) \otimes_t \frac{\partial \mathcal{U}(t,x)}{\partial t} =\frac{\partial \mathcal{U}(t,x)}{\partial t} $, and hence, equation (\ref{eq:5}) reduces to classical KF equation (\ref{eq:4}). The classical KF equation is classical in the sense that when the delta function is used as a memory function, one can trace the exact position and time (memory) of the particle which gives us a classical insight. Now, let us consider that any arbitrary  $m(t)$ memory function has a category of functions $n(t)$ given by the \textbf{condition} $ n(t)\otimes_t m(t)=\delta (t)$ \cite{volatility}, then the equation \eqref{eq:5} can be written in another format as given in \cite{volatility,iaeng} as
\begin{eqnarray}
\label{eq:6}
\tau \frac{\partial \mathcal{U}(t,x)}{\partial t} \ =-n(t) \otimes_t \mathcal{U}(t,x)+ n(t) \otimes_t \mathcal{U}(t,x) \otimes_x f(x) 
\end{eqnarray}
Equation \eqref{eq:6} then becomes the \textit{Generalized Kolmogorov-Feller equation (GKFE)}. Again, if $n(t)=\delta(t)$ in GKFE, then one can easily prove that GKFE yields classical KF equation given by equation \eqref{eq:4}. We will be following the form of GKFE solution as given by Blackledge \textit{et al.} \cite{volatility,iaeng} . After adding $\mathcal{U}(t,x)$ on both sides of equation \eqref{eq:6}and solving using Green function technique, then the GKFE solution is \cite{volatility,iaeng} 
\begin{eqnarray}
\label{eq:v}
\mathcal{U}(t,x)= - g(t) \otimes_t n(t) \otimes_t \mathcal{U}(t,x)+ g(t) \otimes_t n(t) \otimes_t \mathcal{U}(t,x)\otimes_x f(x)+g(t)\otimes_t \mathcal{U}(t,x)
\end{eqnarray}
Here, the function $g(t)$ denotes the Green function which is solution of the following equation,
\begin{eqnarray}
\label{eq:g}
\tau\frac{\partial  g(t-t_0)}{\partial t}+g(t-t_0)=\delta(t-t_0)
\end{eqnarray}
Its solution is $ \ g(t)=\frac{1}{\tau}\mathrm{e}^{({\frac{-t}{\tau}})}$, for time, $t>0$. The proof is simple, and can be proceeded by taking $t-t_0=z$. Then, $g(z)$ is given by $ g(z)= \int\limits_{t_0}^\infty \mathcal{G}(t,z)\ f(z)\ dz$. The $f(z)$ is a known function and $\mathcal{G} (t,z)$ is its fundamental solution. We have,
\begin{eqnarray}
\label{eq:GG} 
\frac{\partial \mathcal{G}(t,z)}{\partial t} +\frac{\mathcal{G}(t,z)}{\tau}=\frac{\delta(t-z)}{\tau}
\end{eqnarray}
Its homogeneous solution is $\mathcal{G}_{hom}(t,z)=\mathrm{e}^{\left (-\int\limits_z^t \ \frac{1}{\tau} \ dt \right )}$. We know that, $\int \limits_{t_0}^\infty \delta(t-z) \ dz=\textbf{H}(t-z)$, where Heaviside function, $\textbf{H}(t-z)$ has the following definition,
\[
 \textbf{H}(t-z)= 
  \begin{cases}
                                   
                                   0 & \text{; $t< z $} \\
                                   1 & \text{; $t> z $} \\
  \end{cases}
\]
Therefore, the fundamental solution is 
$\mathcal{G}(z,t)=\frac{1}{\tau}\mathrm{e}^{\left (-\int\limits_z^t\frac{1}{\tau} \ dt\right)}\times\textbf{H}(t-z)$. Thus, the above solution to equation \eqref{eq:g} is proved to be $g(t)= \frac{1}{\tau} e^{(\frac{-t}{\tau})}$ for time, $t>0 $ with intial $t_0=0$. Now, taking simultaneous Fourier (k-space) and Laplace (s-space) transforms of equation \eqref{eq:v} as discussed in \cite{volatility,iaeng}, one can arrive at,
\begin{eqnarray}
\label{eq:10}
 \widetilde{\overline{\mathcal{U}}}(s,k)=-\overline{g}(s)\ \overline{n}(s)\ \widetilde{\overline{\mathcal{U}}}(s,k)+\overline{g}(s)\ \overline{n}(s)\ \widetilde{\overline{\mathcal{U}}}(s,k)\ \widetilde{f}(k)+\overline{g}(s)\ \widetilde{\overline{\mathcal{U}}}(s,k)
\end{eqnarray}
where, by definition,
\begin{eqnarray} 
\widetilde{\overline{\mathcal{U}}}(s,k)&=&\int \limits_{t=0}^\infty\int\limits_{x=-\infty}^{+\infty} \ e^{-ikx} \ e^{-st} \ \mathcal{U}(t,x) \ dx \ dt \nonumber\\
\overline{g}(s)&=&\int \limits_{t=0}^\infty \ e^{-st} \ g(t)  \ dt;~~
\overline{n}(s)=\int \limits_{t=0}^\infty \ e^{-st}  \ n(t) \ dt;~~
\widetilde{f}(k)=\int \limits_{x=-\infty}^{+\infty} \ e^{-ikx}  \ f(x)\ dx\nonumber
\end{eqnarray}
Now, solving equation \eqref{eq:10}, one can get \cite{volatility}, 
\begin{eqnarray}
\label{eq:uu}
\widetilde{\overline{\mathcal{U}}}(s,k)=-\frac{\overline{g}(s)\ \overline{n}(s)\ \widetilde{\overline{\mathcal{U}}}(s,k)}{[1-\overline{g}(s)]}+ \frac{\overline{g}(s) \ \overline{n}(s) \ \widetilde{\overline{\mathcal{U}}}(s,k) \  \widetilde{f}(k)}{[1-\overline{g}(s)]}
\end{eqnarray}
The functional form of $\overline{g}(s)$ can be obtained by taking Laplace transform of the equation \eqref{eq:g}, where, $y=t$, with the initial condition $g(0)=0$ i.e., one can get as $\tau \left[s\overline{g}(s)-g_0\right] + \overline{g}(s) =1$. From this, one can get the expression, $\overline{g}(s)= \frac{1}{1+ \tau s}$. Now, use $\frac{\overline{g}(s)}{1-\overline{g}(s)}=\frac{1}{\tau s}$ in equation \eqref{eq:uu}, and solve as follows \cite{volatility,iaeng},
\begin{align}
\tau s \ \widetilde{\overline{\mathcal{U}}}(s,k)&=-\overline{n}(s)\ \widetilde{\overline{\mathcal{U}}}(s,k)+ \overline{n}(s)\ \widetilde{\overline{\mathcal{U}}}(s,k)\ \widetilde{f}(k)\nonumber
\end{align}
Solving the equation, we get,
\begin{eqnarray}
\widetilde{\overline{\mathcal{U}}}(s,k)&= \overline{h}(s)\ \widetilde{\overline{\mathcal{U}}}(s,k) \ \widetilde{f}(k) \label{eq:zz}
\end{eqnarray}
where, 
\begin{equation}
\overline{h}(s)=\frac{\overline{n}(s)}{\tau s + \overline{n}(s)} \label{eq:111}
\end{equation}
\noindent Then, one can obtain the functional form of $h(t)$ by the application of Inverse Laplace transform to exoression of $\overline{h}(s)$ of equation \eqref{eq:111}, . This function $h(t)$ relates to the category  memory function $n(t)$ indirectly via equation \eqref{eq:111} and this category function $n(t)$ again is related to the main memory function $m(t)$ via condition \cite{volatility}.\\ \\Further, by applying both Inverse Fourier and Inverse Laplace transforms to $\widetilde{\overline{\mathcal{U}}}(s,k)$ of equation \eqref{eq:zz}, we finally get a self-consistent form of solution $\mathcal{U}(t,x)$ of GKFE as Blackledge \textit{et al.} derived \cite{volatility,iaeng} ,
\begin{eqnarray}
\label{eq:uxt}
\mathcal{U}(t,x)=h(t)\otimes_t \mathcal{U}(t,x) \otimes_x f(x)
\end{eqnarray}
From equation \eqref{eq:uxt}, we can see that the concept of memory is incorporated in the solution of GKFE through $h(t)$. Different functional forms of $h(t)$ will give different GKFE solutions, $ \mathcal{U}(t,x).$ This signifies the role of memory functions in the GKFE solutions. In the subsequent sections, we will study these GKFE solutions with respect to different memory functional forms derived from one particular Mittag-Leffler function. 
\section{Memory functional forms derived from Mittag-Leffler Function, $E_\beta$}
\noindent The Mittag-Leffler (ML) function of order $\beta$, $E_\beta$ represents a complete transcendental function defined in the $\mathbb Z$-complex plane as\ \cite{mainardi,gorenflo,saxena},
\begin{eqnarray}
E_\beta(z):= \sum \limits_{n=0}^\infty \frac{z^n}{\Gamma (\beta n+1)}, \ \ where \ \ \beta >0\nonumber
\end{eqnarray}
Let us denote, for time, $t>0$, 
\begin{eqnarray}
\label{eq:b}
e_\beta (t) := E_\beta(-t^\beta)=\sum\limits_{n=0}^{\infty} \ (-1)^n \ \frac{ t^{\beta n}}{\Gamma (1+\beta n) } \ , \ \ where \ \ \beta \ is\  such \  that \ 0<\beta<1
\end{eqnarray}
A negative exponential function which is characteristic of a memoryless Markovian system is obtained in the limit $\beta \rightarrow 1$. 
\vskip 0.3cm
\subsection{Asymptotic approximations of Mittag-Leffler function}
\noindent From equation (\ref{eq:b}) (the representation of power series which is convergent), the ML function's short time approximation is obtained as follows.
\begin{eqnarray} 
e_\beta^{(0)} (t) = 1- \frac{1}{\Gamma (1+\beta ) }t^{\beta } +...
\sim e^{- \frac{ t^{\beta}}{\Gamma (\beta+1) }}, \ \ \ \  t\rightarrow 0 \label{eq:c}
\end{eqnarray} 
Thus, for small time $ t\rightarrow 0$, the power series ML function of equation \eqref{eq:b} approximates to \textit{stretched exponential} of equation \eqref{eq:c}. Again, from the asymptotic power series representation of equation (\ref{eq:b}), we can get the ML function's long term approximation as \cite{erdel,mainardi,gorenflo},
\begin{eqnarray}
e_\beta^{(\infty)}(t) \sim \sum\limits_{n=1}^{\infty} \ (-1)^{n-1} \ \frac{ t^{-\beta n}}{\Gamma (1- \beta n) }\sim \frac{ t^{-\beta}}{\Gamma (1- \beta) }  \ , \ \ \ \  t \rightarrow \infty \label{eq:d}
\end{eqnarray}
Thus, for large time $ t\rightarrow \infty$, the power series ML function of equation \eqref{eq:b} approximates to \textit{inverse power law} of equation (\ref{eq:d}).
From equations (\ref{eq:c}) and (\ref{eq:d}), we find that the Mittag-Leffler function, $e_\beta (t)$ of equation (\ref{eq:b}) is the interpolation of intermediate time between two limiting behaviours, $t \rightarrow 0$ and $t \rightarrow \infty.$ The stretched exponential of equation \eqref{eq:c} represents a model of the decay which is very fast (or memoryless) for small time. On the otherhand, the inverse power law of equation \eqref{eq:d} is because of the decay very slow in nature (having some memory) for large time. Thus, the Mittag-Leffler function is a good candidate for being a complete memory function since it can represent short term and long term memory in its limits.
The Mittag-Leffler function, $e_\beta(t)$ still maintains its \textquotedblleft \textit{complete monotonic}\textquotedblright\  character \cite{mainardi} for $0<\beta<1$, even though the function loses its behaviour of very fast (exponential) decay by exibiting power law tails at larger time. The \textquotedblleft \textit{complete monotonic}\textquotedblright\  character of the function,  $e_\beta(t)$ means \cite{erdel}   
$$ (-1)^m \ \frac{d^m}{dt^m} e_\beta(t)\  \geq 0,\ \ \ \ \ where \ \ time, t\geq 0, \ \ 0\leq \beta\leq1, \ and  \ \ \  n=0,1,...$$

\section{Results: GKFE solutions driven by different forms of Mittag-Leffler memory function}
\noindent This section presents analytical results regarding different forms of memory functions.  In Case-I, upto first order expansion of  \textit{stretched exponential} of equation \eqref{eq:c} is taken as memory function $m(t).$ In Case-II, the memory function $m(t)$ is taken to be the \textit{negative power law} functional form of equation \eqref{eq:d}. In Case-III, the memory function $m(t)$ is the intermediate functional form of the Mittag-Leffler function of equation \eqref{eq:b}. In Case-IV, negative exponential is taken as a memory function $m(t)$. In all the cases, respective final GKFE solutions are presented. Detailed calculations are given below. \\ \\
\noindent \textbf{Case-I:} \textit{For time, $t \rightarrow 0$ (short memory)} \\ 
\noindent Consider the memory function in this case to be $m(t)=1-\frac{t^\beta}{\Gamma (1+\beta)}$. To find GKFE solution associated with this memory function, we need to find the expression for $\bar{n}(s)$ to get the functional form of $h(t)$. In order to find the form of $\bar{n}(s)$ to get the solution of GKFE, we use the condition $n(t) \otimes_t m(t)= \delta (t)$ \cite{volatility,stochastic} as follows,
\begin{eqnarray}
\label{eq:nn}
\int\limits_0^t n(\tau)\left[1-\frac{(t-\tau)^\beta}{\Gamma (1+\beta)}\right]d\tau =\delta(t)
\end{eqnarray}
Now, applying the change of variables i.e., $t\rightarrow \tau$ and $\tau \rightarrow s$ in equation \eqref{eq:nn} and then multiplying by $(t-\tau)^{-\beta-1}$ and again integrating w.r.t $\tau$, it gives
\begin{eqnarray}
\int\limits_0^t d\tau \ (t-\tau)^{-\beta-1} \int\limits_0^\tau n(s) ds \left [1-\frac{(\tau-s)^\beta}{\Gamma (1+\beta)}\right]&=& \int\limits_0^t d\tau \ \delta (\tau)\ (t-\tau)^{-(\beta+1)}=t^{-(\beta+1)}
\end{eqnarray}
Then, applying Fubini's theorem of change of integrating variables to the L.H.S., we get,  
\begin{eqnarray}
\label{eq:ns}
\int\limits_0^t n(s) ds \int\limits_s^t d\tau \ (t-\tau)^{-\beta-1}\left [1-\frac{(\tau-s)^\beta}{\Gamma (1+\beta)}\right]=t^{-(\beta+1)}
\end{eqnarray}
Next, in order to solve the integration, the change of variables, $Y=\frac{\tau -s}{t-s}$, is introduced where, $\tau= s+Y(t-s)$ and $d\tau=dY (t-s)$. We then integrate the second integration in equation \eqref{eq:ns}, and after simplification and arranging the terms, we arrive at,
\begin{eqnarray}
\int\limits_0^t ds \ n(s) \left[\frac{-(t-s)^{-\beta}}{\beta}-\frac{1}{\Gamma(1+\beta)} \  \textbf{\large{B}}(\beta+1,-\beta) \right]=t^{-\beta-1}
\end{eqnarray}
where, Beta function $ \textbf{\large{B}}(1+\beta,-\beta)=\int\limits_0^1 dY\ Y^\beta (1-Y)^{-1-\beta}$. Using the relation $\textbf{\large{B}}(y,z)=\frac{\Gamma (y) \ \Gamma (z)}{\Gamma(y+z)}$, we get,
\begin{eqnarray} 
\label{eq:sn}
\int\limits_0^t ds n(s)  \left[\frac{-(t-s)^{-\beta}}{\beta}-\Gamma (-\beta)\right]= t^{-(\beta+1)}
\end{eqnarray}
Now, on both sides of the above equation \eqref{eq:sn}, we take Laplace transform and after simplification, we get,
\begin{eqnarray}
\label{eq:ln}
\bar{n}(s)\left[\frac{-L\left \{t^{-\beta}\}\right]}{\beta}-\Gamma (-\beta)L\left\{1\right\}\right]= L\{t^{-(\beta+1)}\}
\end{eqnarray}
By the definition of Laplace transform, we know, $L\left \{t^{-\beta}\right\}=\frac{\Gamma (1-\beta)}{s^{1-\beta}}$, $L\left\{1\right\}=1/s$, and $L\{t^{-\beta-1}\}=s^{\beta} \  \Gamma (-\beta) $. Then using the property, $ \Gamma(1-\beta)=-\beta \ \Gamma (-\beta)$, we obtain the expression for $\bar{n}(s)$ as follows,
\begin{eqnarray}
\label{eq:bn}
\bar{n}(s)=\frac{s}{1-s^{-\beta}}
\end{eqnarray}
Now, the expression for $\bar{n}(s)$ is put in the equation \eqref{eq:111}, and for the condition $\tau \rightarrow 0$, it can be shown that $\bar{h}(s)=1$. Then taking inverse Laplace transform, we can obtain the expression for $h(t)$ as, $h(t)\approx\delta(t)$,  which is characteristic of classical Markovian process. Then putting the expression for $h(t)$ in the equation \eqref{eq:uxt}, we obtain the self-consistent GKFE solution for this short term memory function as given by,
\begin{eqnarray}
\label{eq:stm}
\mathcal{U}(t,x)&=& \mathcal{U}(t,x) \otimes_x f(x)
\end{eqnarray}
Iteratively, this solution can be represented as $\mathcal{U}_{n+1}(t,x)= \mathcal{U}_n(t,x) \otimes_x f(x)$, with convolution over $x$. Here, the function $f(x)$ can be any known function according to one's choice. f(x) is system dependent and one can analyse the resultant equation \eqref{eq:stm} according to it. For example, let us take a Gaussian distribution given by, $f(x)=\frac{1}{\sqrt{2\pi}}e^{-\frac{x^2}{2}}$. Hence, equation \eqref{eq:stm} becomes, $\mathcal{U}(t,x)=\mathcal{U}(t,x) \otimes_x \frac{1}{\sqrt{2\pi}}e^{-\frac{x^2}{2}}.$    \\ \\
\noindent \textbf{Case-II:} \textit{For large time, $t\rightarrow \infty$ (long memory)}\\ 
\noindent Consider memory function to be $m(t)=\frac{t^{-\beta}}{\Gamma (1-\beta)}$. In this case also, in order to find out the form of $\bar{n}(s)$, we use the same condition $n(t) \otimes_t m(t)= \delta (t)$,
\begin{eqnarray}
\label{eq:ns2}
\int\limits_0^t n(\tau) \frac{d\tau}{\Gamma(1-\beta)(t-\tau)^\beta}=\delta(t)
\end{eqnarray}
We now replace $(1-\beta)=\alpha$ in the equation \eqref{eq:ns2}. We proceed the change of the variable from $t$ to $\tau$ and $\tau$ to $s$. Further, in the equation, the term $(t-\tau)^{-\alpha}$ is multiplied on both sides and with reference to $\tau$, it is integrated. Then, the other of the integration is exchanged using Fubini's theorem and after rearranging the terms, we get,
\begin{eqnarray}
\label{eq:ns3}
\int\limits_0^t n(s)\ ds \ \int \limits_s^t\frac{1}{(t-\tau)^\alpha}\times \frac{1}{ (\tau-s)^{1-\alpha}}d\tau=\int \limits_0^t \frac{\Gamma (\alpha) \ \delta(\tau)\ d\tau}{(t-\tau)^{\alpha}}
\end{eqnarray}
Now, we put, $\tau=s+Y(t-s)$ in equation \eqref{eq:ns3}, such that $d\tau=dY (t-s)$ in L.H.S. We then use the Beta function property, $ \int \limits_0^1 Y^{\alpha-1} (1-Y)^{-\alpha} dY =\textbf{B}(\alpha, 1-\alpha)=\Gamma(\alpha) \Gamma(1-\alpha)$, we get,
\begin{eqnarray}
\label{eq:ns4}
\int\limits_0^t \ \Gamma(1-\alpha)  \ n(s) \ ds =\int \limits_0^t \frac{1}{(t-\tau)^{\alpha}}\delta(\tau)\ d\tau
\end{eqnarray}
From this, we can get the expression for $n(t)$, given by,
\begin{eqnarray}
n(t)&=&\frac{1}{\Gamma(1-\alpha)}\times \frac{d}{dt} \int \limits_0^t \frac{\delta(\tau)\ d\tau}{(t-\tau)^{\alpha}}=\frac{1}{\Gamma(\beta)} \frac{d}{dt} t^{-\alpha}=\frac{1}{\Gamma(\beta-1) \ t^{2-\beta}}
\end{eqnarray}
Using the standard integral, $\int\limits_0^\infty \frac{1}{\Gamma (\beta) \ t^{1-\beta}} \exp(-st) dt=\frac{1}{s^\beta}$, we get, $\overline{n}(s)= \frac{1}{s^{\beta -1}}$. Now, by substituting this form of $\overline{n}(s)$ to the equation \eqref{eq:111}, we get $\overline{h}(s)=\frac{1}{1+\tau s^\beta}$. If we consider a non-Markovian process i.e.,$\tau \gg$ 1, then $\overline{h}(s)=\frac{1}{\tau s^\beta} $ and taking inverse Laplace transform of $\overline{h}(s)$, we can write,
$h(t)\sim \frac{1}{\tau \ \Gamma (\beta) \ t^{1-\beta}}$. Thus, the GKFE solution, which incorporates long time memory is,
\begin{eqnarray}
\label{eq:lpo}
\mathcal{U}(t,x)&=& \frac{1}{\tau \ \Gamma (\beta) \ t^{1-\beta}} \otimes_t \mathcal{U}(t,x) \otimes_x f(x)
\end{eqnarray}
Iteratively, this solution can also be represented as $\mathcal{U}_{n+1}(t,x)= \frac{1}{\tau \ \Gamma (\beta) \ t^{1-\beta}} \otimes_t \mathcal{U}_n(t,x) \otimes_x f(x)$, with convolution over $t$ and $x$. We observe that this final GKFE solution for long term memory depends on the parameter, $\beta.$ Here, the function $f(x)$ is system dependent and any known function can be taken according to one's choice for analysis. For example, let us consider a Gaussian distribution, $f(x)=\frac{1}{\sqrt{2\pi}}e^{-\frac{x^2}{2}}$. Hence, equation \eqref{eq:lpo} becomes, $\mathcal{U}(t,x)=\frac{1}{\tau \ \Gamma (\beta) \ t^{1-\beta}} \otimes_t \mathcal{U}(t,x) \otimes_x \frac{1}{\sqrt{2\pi}}e^{-\frac{x^2}{2}}.$    \\ \\
\noindent \textbf{Case-III:} \textit{For intermediate time, $0<t<\infty$}\par
\noindent In this case, the memory function is taken to be the Mittag-Leffler function or, $m(t)=e_\beta (t)=E_\beta (-t^{\beta})$, given by equation({\ref{eq:b}}). We then substitute this function directly to the relation $n(t) \otimes_t m(t)= \delta (t)$ \cite{volatility} such that $n(t) \otimes_t E_\beta(-t^{\beta})=\delta (t)$. Taking on both sides of this equation the Laplace transform, and employing the relation $L\{E_\beta (-t^{\beta})\}=\frac{s^{\beta-1}}{1+s^\beta}$ \cite{saxena,feller}, we get the expression for $\bar{n}(s)$ given by, $\bar{n}(s)=\frac{1+s^\beta}{s^{\beta-1}}$. Now, putting this expression of $\bar{n}(s)$ to the equation \eqref{eq:111}, we get the expression for $\bar{h}(s)$ given by, $\bar{h} (s)=\frac{1+s^\beta}{1+s^\beta+\tau s^\beta}$ and $h(t)$ will be its inverse Laplace transform which also depends on $\beta$. The form of the GKFE solution in this case will be given by $h(t) \otimes_t \mathcal{U}(t,x) \otimes_x f(x).$ Iteratively, this solution can also be represented as $\mathcal{U}_{n+1}(t,x)= h(t)\otimes_t \mathcal{U}_n(t,x) \otimes_x f(x)$, with convolution over $x$ and $t$ and depends on $\beta $. Here, the function $f(x)$ is system dependent and any known function can be taken according to one's choice for analysis. For example, let us consider a Gaussian distribution, $f(x)=\frac{1}{\sqrt{2\pi}}e^{-\frac{x^2}{2}}$. Hence, $\mathcal{U}(t,x)=h(t) \otimes_t \mathcal{U}(t,x) \otimes_x \frac{1}{\sqrt{2\pi}}e^{-\frac{x^2}{2}}.$   \\
\\
\noindent \textbf{Case-IV:} For $\beta\rightarrow 1$, using equation \eqref{eq:b}, the Mittag-Leffler function approximates to,
\begin{eqnarray}
e(t)_{(\beta\rightarrow 1)}  =\sum\limits_{n=0}^{\infty} \  \ \frac{ (-1)^n \ t^{ n}}{\Gamma (n+1)} =e^{-t}  \nonumber
\end{eqnarray}
Now, consider the memory function to be exponentially decaying $m(t)=e^{-t}$. To find the GKFE solution for this memory function, we use the condition $n(t) \otimes_t m(t)= \delta (t)$ \cite{volatility,stochastic} such that after taking Laplace transform on both sides of the condition, 
$L[n(t)] \ L[e^{-t}]=1$, such that, $\bar{n}(s) \ \frac{1}{1+s} = 1$, we get, $\bar{n}(s)=1+s$. Now, equation \eqref{eq:111} becomes,   
\begin{eqnarray}
\bar{h}(s)=\frac{1+s}{\tau s+(1+s)}=\left(1+\frac{\tau s}{1+s}\right)^{-1}=1-\frac{\tau}{1+\frac{1}{s}}=1-\tau +\frac{\tau}{s}\nonumber 
\end{eqnarray}
For memoryless process $(\tau \rightarrow 0)$, $\bar{h}(s)=1$ and, $h(t)=\delta (t).$ Hence, from equation \eqref{eq:uxt}, the self-consistent GKFE solution for this memory function becomes, 
\begin{equation}
\mathcal{U}(t,x)= \delta (t)\otimes_t \mathcal{U}(t,x) \otimes_x f(x) =\mathcal{U}(t,x) \otimes_x f(x) \label{eq:number}    
\end{equation}
Iteratively, this solution can be represented as $\mathcal{U}_{n+1}(t,x)= \mathcal{U}_n(t,x) \otimes_x f(x)$, with convolution over $x$. Here, the function $f(x)$ can be any known p.d.f according to system. For example, let us take a Gaussian distribution given by, $f(x)=f(x|\mu=0,\sigma^2=1)=\frac{1}{\sqrt{2\pi}}e^{-\frac{x^2}{2}}$. Hence, $\mathcal{U}(t,x)=\mathcal{U}(t,x) \otimes_x \frac{1}{\sqrt{2\pi}}e^{-\frac{x^2}{2}}.$ \\ \\
\section{Discussions}
\noindent \textbf{Theorem 1:} The Generalised Kolmogorov-Feller equation solutions are the same for negative exponential function and for upto first order expansion of stretched exponential function for very small $\tau$.\\ \\
\textbf{Proof:} In Case-IV, for negative exponential memory function $m(t)=e^{-t}$, $\bar{h}(s)=  1-\tau +\frac{\tau}{s}.$ For very small $\tau$ or, $(\tau \rightarrow 0)$, $\bar{h}(s)=1 , or \ h(t)=\delta (t).$ Now, in case-I, for first order expansion of stretched exponential function as memory function i.e., $m(t)=1-\frac{t^{\beta}}{\Gamma(\beta +1)}$, we get, $\bar{h}(s)=\frac{1}{1+\tau-\frac{\tau}{s^\beta}}.$ Again, for $(\tau \rightarrow 0), \ \bar{h}(s)=1 , or \ h(t)=\delta (t).$ From equation, \eqref{eq:uxt}, the form of Generalised Kolmogorov Feller Equation solution will be given by $\mathcal{U}(t,x)=h(t)\otimes_t \mathcal{U}(t,x)\otimes_x f(x).$ Hence, in both the cases, the final solution is $\mathcal{U}(t,x)=\delta(t)\otimes_t \mathcal{U}(t,x)\otimes_x f(x)=\mathcal{U}(t,x)\otimes_x f(x) $, for any arbitrary p.d.f $f(x)$, where we have used the property of $\delta$-function that $\delta (t)\otimes_t f(t)=f(t)$. Hence proved. \\ \\
\noindent\textbf{Theorem 2:} \textit{Mittag-Leffler function for intermediate memory function cross-overs from $t\rightarrow 0$ to $t\rightarrow \infty$.}\\ \\
\noindent \textbf{Proof:}\textit{
The expression of $\bar{n}(s)$ for intermediate memory function $(0<t<\infty)$ is $\bar{n}(s)=\frac{1+s^\beta}{s^{\beta-1}}$. The expression for $\bar{h}(s)$ is obtained by substituting the expression of $\bar{n}(s)$ to the equation \eqref{eq:111}, and is given by, $\bar{h} (s)=\frac{1+s^\beta}{1+s^\beta+\tau s^\beta}$.
Now, if $\tau \ll 1$, this equation for $\bar{h}(s)$ becomes $\bar{h}(s)\approx1$. Taking its Inverse Laplace transform, we get the same result which was obtained in the Case-I, $h(t)\approx\delta(t)$. On the other hand, rewriting the expression for $\bar{h}(s)$ in the form, $\bar{h}(s)=\frac{1+s^\beta}{1+s^\beta+\tau s^\beta}=\frac{1}{1+\frac{s^\beta \tau}{1+s^\beta}}=\frac{1}{1+\frac{\tau}{1+s^{-\beta}}}$, and, if $\tau\gg 1$, then the expression of $\bar{h}(s)$ becomes the same expression which we got in the Case-II, $\bar{h} (s)\approx\frac{1}{\frac{\tau}{1+s^{-\beta}}}\approx\frac{1+s^{-\beta}}{\tau}\approx\frac{s^{-\beta}}{\tau}+\frac{1}{\tau}\approx\frac{1}{\tau s^\beta}$. Taking its Inverse Laplace transform, we get the same result which was obtained in the Case-II, $h(t)\sim \frac{1}{\tau \ \Gamma (\beta) \ t^{1-\beta}}$. Thus, we can conclude the Mittag-Leffler function in intermediate case provides us the cross-over from short to long term memory results as discussed in the Case-I and Case-II.} \\ \\
\noindent \textbf{Theorem 3:} \textit{A generalized integro-differential equation of the GKFE is,}
\begin{eqnarray}
\label{eq:GKFE}
\frac{\partial}{\partial t} \left[ \int\limits_0^x \int\limits_0^t \ dx_1 \ dt_1 \ h(t_1) \ \mathcal{U}(x_1,t-t_1) \ f(x-x_1) \right]
&=&-\frac{1}{\tau }\int \limits_0^x \int \limits_0^t \ dx_1 \ dt_1 \ n(t_1)\ \mathcal{U}(x_1,t-t_1) \ f(x-x_1) \nonumber\\
&&+\frac{1}{\tau }\int \limits_0^x  \int \limits_0^x \int \limits_0^t dx_2 dx_1 dt \ n(t_1) \ \mathcal{U}(x_2,t-t_1)\ f(x_1) f(x-x_2-x_1)\nonumber \\
\end{eqnarray}
\noindent\textbf{Proof:}\textit{
In order to write a generalized integro-differential equation of the GKFE of equation (\ref{eq:6}), we can write the left hand side of the equation \eqref{eq:6},
\begin{eqnarray}
\tau \frac{\partial \mathcal{U}(t,x)}{\partial t }&=&\tau \frac{\partial}{\partial t } \left [ h(t)\otimes_t \mathcal{U}(t,x) \otimes_x f(x)\right ]=\tau \frac{\partial}{\partial t} \left[ h(t)\otimes_t \int\limits_0^x \mathcal{U}(t,x_1) \ f(x-x_1) \ dx_1\right]\nonumber\\
&=& \tau \frac{\partial}{\partial t} \left[ \int\limits_0^x \int\limits_0^t \ dx_1 \ dt_1 \ h(t_1) \ \mathcal{U}(x_1,t-t_1) \ f(x-x_1) \right]\nonumber
\end{eqnarray}
Again, the right hand side of the equation \eqref{eq:6} becomes,
\begin{eqnarray}
&&-n(t)\otimes_t \mathcal{U}(x,t) \otimes_x f(x)+n(t)\otimes_t \mathcal{U}(x,t) \otimes_x f(x) \otimes_x f(x)\nonumber\\
&=&-n(t)\otimes_t \int \limits_0^x \mathcal{U}(x_1,t) \ f(x-x_1) \ dx_1+ n(t) \otimes_t \mathcal{U}(x,t) \otimes_x \int \limits_0^x f(x_1) \  f(x-x_1) dx_1\nonumber\\ 
&=&-\int \limits_0^x n(t) \otimes_t \mathcal{U}(x_1,t) \ f(x-x_1) \ dx_1+ n(t)\otimes_t \int \limits_0^x \mathcal{U}(x_2,t) \int \limits_0^x f(x_1) \ f(x-x_2-x_1) dx_2 dx_1\nonumber\\
&=&-\int \limits_0^x \int \limits_0^t n(t_1) \  \mathcal{U}(x_1,t-t_1) \ dt_1\ f(x-x_1) \ dx_1+ \int \limits_0^t n(t_1) \int \limits_0^x \mathcal{U}(x_2,t-t_1) \  dt_1 \int \limits_0^x f(x_1) \ f(x-x_2-x_1) \ dx_2 \  dx_1\nonumber\\
&=&-\int \limits_0^x \int \limits_0^t \ dx_1 \ dt_1 \ n(t_1)\ \mathcal{U}(x_1,t-t_1) \ f(x-x_1) +  \int \limits_0^x  \int \limits_0^x \int \limits_0^t n(t_1) \ \mathcal{U}(x_2,t-t_1)\ f(x_1)\ f(x-x_2-x_1)\nonumber
\end{eqnarray}
Hence, the generalized integro-differential equation of the GKFE in equation \eqref{eq:GKFE} is proved.}\\ \\
\noindent \textbf{Lemma 1:} \textit{The integro-differential form of GKFE for the asymptotic case (i.e., $t\rightarrow \infty$) is}
\begin{eqnarray}
\label{eq:inf}
\frac{1}{\beta} \frac{\partial}{\partial t} \left[ \int\limits_0^x \int\limits_0^t \ dx_1 \ dt_1 \ t_1^{\beta-1} \ \mathcal{U}(x_1,t-t_1) \ f(x-x_1) \right]
&=&-\frac{\Gamma(\beta)}{\tau \ \Gamma(2\beta-1)}\left[\int \limits_0^x \int \limits_0^t dx_1 \ dt_1 \ t_1^{2\beta-2} \ \mathcal{U}(x_1,t-t_1) \ f(x-x_1)\right.\nonumber\\
&&+\left. \int \limits_0^x \int \limits_0^x \int \limits_0^t dx_2dx_1dt_1 \ t_1^{2\beta-2} \  \mathcal{U}(x_2,t_1-t) f(x_1) f(x-x_2-x_1)\right] \nonumber \\
\end{eqnarray}
\noindent \textbf{Proof:}\textit{
For the case of large time, $t\rightarrow \infty$, we get,
\begin{eqnarray}
\mathcal{U}(x,t)= \frac{1}{\tau \Gamma (\beta)}t^{\beta-1} \otimes_t \mathcal{U}(x,t) \otimes_x f(x)\nonumber
\end{eqnarray} 
Putting this to the equation (\ref{eq:6}), L.H.S. of equation \eqref{eq:6} becomes,
\begin{eqnarray}
\tau \frac{\partial}{\partial t } \mathcal{U}(t,x)=\tau \frac{\partial}{\partial t } \left [ \frac{t^{\beta-1}}{\tau \ \Gamma (\beta)} \otimes_t \mathcal{U}(x,t) \otimes_x f(x)\right ]=\frac{1}{\Gamma(\beta)} \frac{\partial}{\partial t} \left[ \int\limits_0^x \int\limits_0^t \ dx_1 \ dt_1 \ t_1^{\beta-1} \ \mathcal{U}(x_1,t-t_1 \ f(x-x_1) \right]\nonumber
\end{eqnarray}
\textit{For $n(t)=\frac{t^{\beta-2}}{\Gamma(\beta-1)}$, we get,}
\begin{eqnarray}
R.H.S.~of~(6)&=&-\frac{t^{\beta-2}}{\Gamma(\beta-1)}\otimes_t \frac{t^{\beta-1}}{\tau \ \Gamma (\beta)} \otimes_t \mathcal{U}(x,t) \otimes_x f(x)+\frac{t^{\beta-2}}{\Gamma(\beta-1)}\otimes_t \frac{t^{\beta-1}}{\tau \ \Gamma (\beta)} \otimes_t \mathcal{U}(x,t) \otimes_x f(x)\otimes_x f(x)\nonumber
\end{eqnarray} 
Now, 
\begin{eqnarray}
t^{\beta-2}\otimes_t t^{\beta-1}&=&\int \limits_0^t \tau^{\beta-2}\ (t-\tau)^{\beta-1}\ d\tau = \int \limits_0^t \tau^{\beta-2} \ \ t^{\beta-1}\ \left(1-\frac{\tau}{t}\right)^{\beta-1} \ d\tau \nonumber\\
&=& \int \limits_0^1 (tX)^{\beta-2} \ \ t^{\beta-1}\ \left(1-X\right)^{\beta-1} \ t \ dX \ \ \ \ \ \ \ \ \ \ \ \ \ (Putting \ \tau=tX) \nonumber\\
&=&t^{2\beta-2}\int \limits_0^1  \ \ \left(1-X\right)^{\beta-1} \ X^{\beta-1-1}\ dX = t^{2\beta-2} \ \textbf{B}(\beta-1,\beta)= \frac{t^{2\beta-2} \  \Gamma(\beta-1) \ \Gamma (\beta)}{\Gamma(2\beta-1)}\nonumber
\end{eqnarray}
\begin{eqnarray}
\therefore R.H.S.~of~(6) &=&-\frac{\Gamma(\beta)}{\tau \ \Gamma(2\beta-1)}\left[t^{2\beta-2} \otimes_t \mathcal{U}(x,t)\otimes_x f(x)+ t^{2\beta-2} \otimes_t \mathcal{U}(x,t) \otimes_x f(x)\otimes_x f(x)\right]\nonumber\\
&=&-\frac{\Gamma(\beta)}{\tau \ \Gamma(2\beta-1)}\left[\int \limits_0^x \int \limits_0^t dx_1 \ dt_1 \ t_1^{2\beta-2} \  \mathcal{U}(x_1,t-t_1) \ f(x-x_1)\right.\nonumber\\
&&\left.+\int \limits_0^x \int \limits_0^x \int \limits_0^t dx_2\ dx_1 \ dt_1\ t_1^{2\beta-2}   \  \mathcal{U}(x_2,t_1-t) \ f(x_1) \ f(x-x_2-x_1)\right]\nonumber
\end{eqnarray}
\textit{Hence, the generalized integro-differential equation of the GKFE in equation \eqref{eq:inf} is proven.}}
\section{Conclusion}
\noindent In conclusion, we have found the self-consistent solutions of the GKFE for four cases i.e., when the memory functions are Mittag-Leffler function, derived negative exponential, and ML function's approximations at short time and long time. We have found that the GKFE solutions are the same for negative exponential and for upto frst order expansion of stretched exponential function for very small $\tau \rightarrow 0$. The cross-over from short time to long time approximations of the ML function in GKFE is presented in this paper. We have also found a generalized integro-differential equation of the GKFE and have expressed the same for an asymptotic case. We conclude by saying that modelling complex dynamics of natural systems with non-Markovian GKFE can make the complex systems study more realistic in nature by considering past event histories. One important area of our study can be in modelling short term and long term memory in neuroscience. We would like to add that the GKFE with memory function is a very important formalism which can give valuable insights about stochastic volatility or extreme events in many complex systems study such as complex biological systems (like disease mechanisms in epilepsy); chemical systems such as important information processing properties of complex molecular events driven at short time and at a longer time; financial system such as stock market and also social systems such as outbreak of an epidemic. \\
 
\noindent\textbf{Authors Contribution}\\
\noindent The present work is conceptualized by RKBS and ALC. ALC and JB produced analytical results. The manuscript is written by ALC and RKBS. All authors discussed, finalised and approved the presented manuscript.\\

\noindent\textbf{Competing financial interests:}\\
\noindent The authors declare no competing financial interests.

\end{document}